\documentclass{article} 
\usepackage{iclr2026_preprint,times}

\usepackage{amsmath,amsfonts,bm}

\def\eqref#1{equation~\ref{#1}}

\def\1{\bm{1}}

\DeclareMathAlphabet{\mathsfit}{\encodingdefault}{\sfdefault}{m}{sl}
\SetMathAlphabet{\mathsfit}{bold}{\encodingdefault}{\sfdefault}{bx}{n}

\usepackage{hyperref}       
\usepackage{url}            
\usepackage{booktabs}       
\usepackage{amsfonts}       
\usepackage{nicefrac}       
\usepackage{microtype}      
\usepackage{xcolor}
\usepackage{csquotes}
\usepackage{enumitem}
\usepackage{graphicx}
\usepackage{amsmath}
\usepackage{color, soul}    
\usepackage{caption}        
\usepackage{multirow, multicol}
\usepackage{wrapfig}
\usepackage{hyperref}
\usepackage{url}

\newcommand{\name}{VLTL-Bench}
\newcommand{\fullname}{Verifiable Linear Temporal Logic Benchmark}

\title{Verifiable Natural Language to Linear Temporal Logic Translation: A Benchmark Dataset and Evaluation Suite}

\author{%
    William English\textsuperscript{1}\thanks{\textsuperscript{1} University of Florida, Gainesville, FL, USA,  \textsuperscript{2} Florida International University, Miami, FL, USA}\\
    \texttt{will.english@ufl.edu} \\
    \And
    Chase Walker\textsuperscript{1} \\
    \And
    Dominic Simon\textsuperscript{1} \\
    \And
    Sumit Jha\textsuperscript{2} \\
    \And
    Rickard Ewetz\textsuperscript{1} \\
    \texttt{rewetz@ufl.edu} \\
}
\iclrfinalcopy 

\begin{document}

\maketitle

\begin{abstract}
Empirical evaluation of state-of-the-art natural language (NL) to temporal logic (TL) translation systems reveals near-perfect performance on existing benchmarks. However, current studies only measure the accuracy of the \emph{translation} of NL logic into formal TL, ignoring a system’s capacity to \emph{ground} atomic propositions into new scenarios or environments. This is a critical feature, necessary for the \emph{verification} of resulting formulas in a concrete state space. 
In this paper, we introduce the \textbf{\fullname{} (\name)}, a unifying benchmark for automated NL-to-LTL translation. 
The dataset consists of three unique state spaces and thousands of diverse natural language specifications and their corresponding formal temporal logic specifications. Moreover, the benchmark contains sample traces to verify the temporal logic expressions. While the benchmark directly supports end-to-end evaluation, we observe that many frameworks decompose the process into i) lifting,  ii) grounding, iii) translation, and iv) verification. The benchmark provides ground truths after each of these steps to enable researchers to improve and evaluate different substeps of the overall problem.  
 Using the benchmark, we evaluate several state‑of‑the‑art NL-to-TL translation models and frameworks, including nl2spec, NL2TL, NL2LTL, Lang2LTL, sequence-to-sequence translation, and various LLM prompting techniques. 
 Our evaluation confirms that existing work is capable of reliably performing lifting and translation with high accuracy, while it exposes their struggles to ground the translation into a state space, which stems from the lack of existing datasets. 
 

\end{abstract}

\section{Introduction}
\label{sec:intro}

Formal verification is essential for the safe deployment of autonomous robots \citep{RobotLanguage, Spec2}, cyber-physical controllers \citep{SurveyTemporal}, and safety-critical software systems \citep{SAS_textbook_PoCPS}. Verification first begins with a specification that defines intent in precise temporal logic (TL) \citep{watson2005autonomous, bellini2000temporal}. However, human stakeholders typically articulate intent in ambiguous natural language (NL) \citep{veizaga2021systematically, lamar2009linguistic, lafi2021eliciting}, and the conversion of this NL to TL is a challenging and time-consuming process that requires human experts \citep{yin2024formal, cardoso2021review, thistle1986control}. Due to this complexity, automated NL-to-TL translation has emerged as a core research problem \citep{NL2TL, Natural2CTL_data, circuits_data, conformalLTL_data}. Recently, neural sequence-to-sequence models \citep{hahn2022formal, pan2023data, hsiung2022generalizing}, grammar-constrained decoders \citep{post2018fastlexicallyconstraineddecoding, geng2024grammarconstraineddecodingstructurednlp}, and large language models (LLMs) \citep{xu2024learning, NL2TL, NL2LTL, nl2spec} have all demonstrated promising results on benchmark corpora, with reported accuracies often exceeding $90\%$.  

Despite these gains, evaluations are misleading as most datasets only test \emph{lifted} translation, where temporal logic formulas contain abstract placeholders for atomic propositions (APs). The harder task of \emph{grounded} translation---instantiating APs with domain-specific actions and arguments---is usually left unmeasured. This imbalance stems from limitations of current datasets, which omit the annotations required to separately evaluate lifting, translation, and grounding. As a result, current frameworks optimize for partial tasks, leaving open the more difficult but necessary problem of grounding for producing fully executable specifications.  

\begin{figure}[!t]
    \centering
    \includegraphics[width=1.0\linewidth]{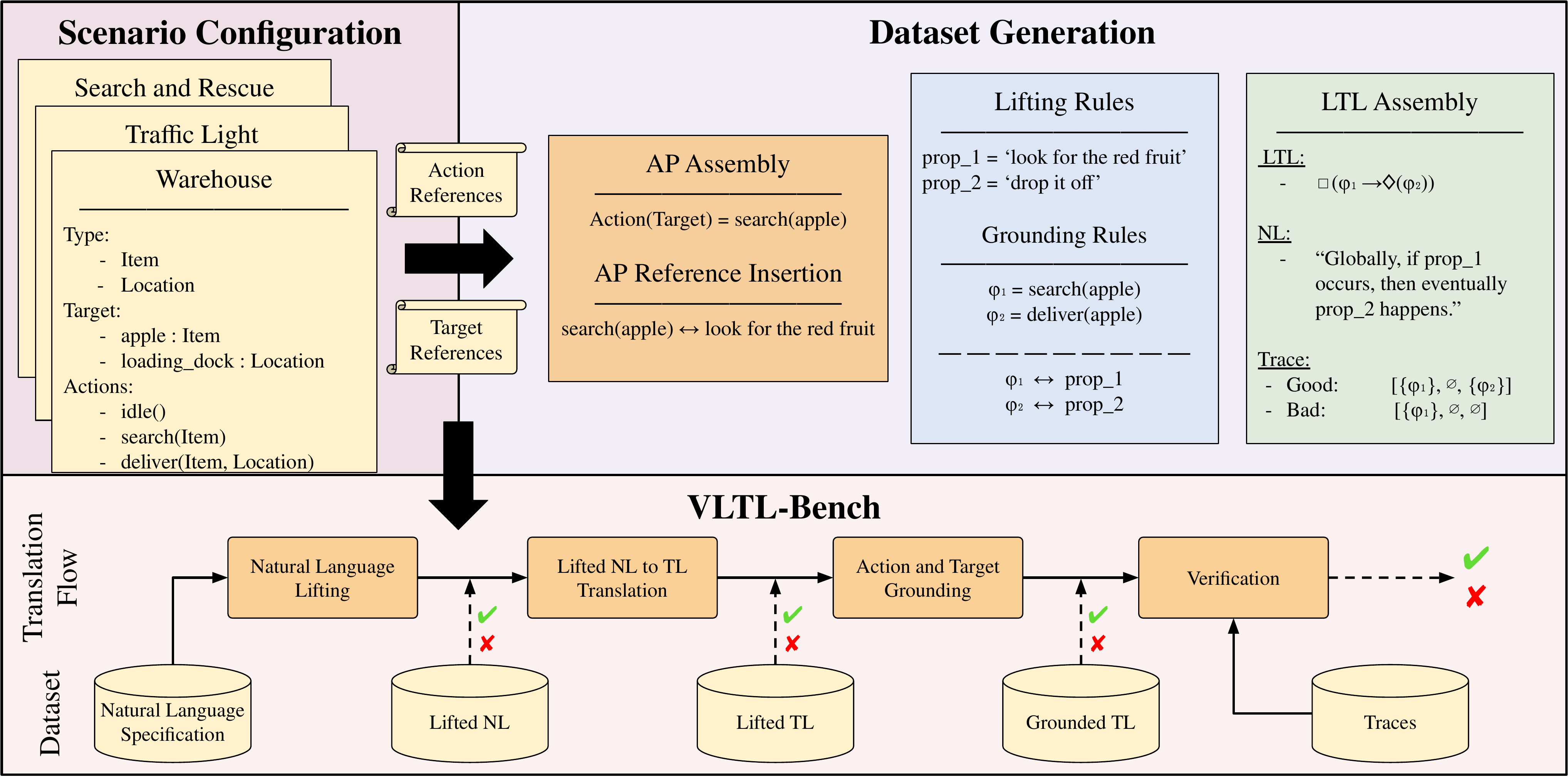}
    \caption{Overview of our dataset synthesis and evaluation framework for NL-to-LTL translation. 
    The framework used a configuration file to define concrete and unique scenarios. The data synthesis generates the NL and TL pairs with associated traces for verification while providing ground truth results for intermediate components. 
} 
    \label{fig:overview}
    \vspace{-0.5cm}
\end{figure}

Benchmarks for NL-to-TL translation include CW \citep{CW_data}, GLTL \citep{GLTL-data}, Navi \citep{Navi_data}, and Conformal \citep{conformalLTL_data}. Their limitations are fourfold. (i) Although recent frameworks decompose the task into lifting, translation, and grounding, these benchmarks supply ground truth only for the end-to-end result (NL-TL pairs), preventing assessment of intermediate components. (ii) CW and GLTL omit grounding entirely, yielding translations without executable semantics. For example, the NL specification: ``Go to the green room and then go to the blue room." is mapped to the LTL expression: ``$\diamondsuit G \rightarrow \diamondsuit B$", without providing a grounded definition of the predicates $G$ and $B$. (iii) Navi and Conformal nominally support grounding but rely on overly simplistic state spaces (e.g., Navi’s colored-room grid), which fails to capture the referential and contextual ambiguities of natural language. (iv) Execution traces/trajectories for independent semantic verification (e.g., via model checking), are not provided, preventing rigorous evaluation.




In this paper, we introduce the \textbf{\fullname{} (\name{})}, a benchmark that grounds linear temporal logic (LTL) in a concrete world state space while broadening linguistic and logical coverage through more diverse atomic propositions. As illustrated in Figure~\ref{fig:overview}, \name{} exposes every stage of the NL-to-TL pipeline: raw and lifted NL specifications, an AP-to-Reference dictionary, lifted and grounded LTL formulas, and both satisfying and unsatisfying traces. Our dataset synthesis and evaluation framework for NL-to-LTL translation leverages scenario configurations to construct grounded action/target combinations, from which we synthesize diverse natural language representations and integrate them into sentence, LTL, and trace templates, yielding corpora whose components can be used individually or combined for holistic evaluation. This layered design makes it possible to isolate performance on lifting, translation, grounding, and verification individually, while also enabling end-to-end evaluation. \
We provide \textcolor{black}{four} scenario configuration files and construct a \textcolor{black}{Kitchen Assistant}, Traffic light, Search \& Rescue, and Warehouse dataset. Using these \textcolor{black}{fou}r datasets we evaluate the capabilities and limitations of state-of-the-art NL to TL translation frameworks.
In summary, we propose: (i) a single, extensible benchmark for evaluating all NL-to-TL translation components; (ii) the first verification evaluation using satisfying and unsatisfying traces; and (iii) an empirical study that reveals both new failure modes in current methods and the severe accuracy decline when grounding is required. 

The remainder of this paper is organized as follows. Section \ref{sec:related} covers preliminaries for LTL and model checking. Section \ref{sec:method} contains a detailed description of the \fullname{} datasets, as well as details on how they were synthesized. Section \ref{sec:experiments} includes an evaluation of current NL-to-TL frameworks on both \fullname{} and existing datasets. We conclude our paper in \ref{sec:conclusion}. Additional details may be found in the Appendix \ref{app}




\section{Background and Related Work}
\label{sec:related}
In this section we introduce necessary notation and background information on temporal logic systems including terminology, linear temporal logic symbols, and existing NL-to-TL datasets. 

\textbf{Linear Temporal Logic. }
The syntax of LTL is given by the following grammar:
\begin{align*}
    \varphi ::={} & \pi 
    \;\mid\; \neg \varphi 
    \;\mid\; \varphi_1 \wedge \varphi_2 
    \;\mid\; \varphi_1 \vee \varphi_2 
    \;\mid\; \varphi_1 \Rightarrow \varphi_2 \nonumber\\
    &\;\mid\; \bigcirc \varphi 
    \;\mid\; \diamondsuit \varphi 
    \;\mid\; \Box \varphi 
    \;\mid\; \varphi_1 \,\cup\, \varphi_2
    \label{eq:ltl-syntax}
\end{align*}
We further discuss model checking with linear temporal logic in Appendix \ref{appendix:ltl} and Appendix \ref{appendix:modelcheck}.

\subsection{Preliminaries}
In this section, we formally define a number of key terms necessary to describe and evaluate NL-to-TL translation systems. 
In order to provide a cogent description of these systems, as well as a robust evaluation, we define these terms as follows:
\begin{enumerate}[wide, labelwidth=!, labelindent=-5pt]
    \item[]\label{prelim:scenario} \textbf{Scenario:} Referred to in existing work as the ``World", ``Environment", or ``Space". A set $S$ of conditions appearing on a trace. 
    \item[] \textbf{Condition:} In model checking, a condition is a uniquely-named Boolean variable $c_i$.   
    \item[]\label{prelim:ap}  \textbf{Atomic Proposition:} $\pi \in \Phi$, where $\Phi$ is the set of propositional variables in an LTL expression. During LTL verification, $\pi_i$ is assigned a value by matching with a condition $c \in S$.
    \item[]\label{prelim:lifting}  \textbf{Lifting:} $\lambda$: NL $\rightarrow \Phi$, extracting substrings corresponding to APs from natural language.
    \item[]\label{prelim:grounding}  \textbf{Grounding:} $g(\pi) = c$, replacing an abstract AP in an LTL expression with a condition $c \in S$.
    \item[]\label{prelim:translation}  \textbf{Translation:} $\tau$: NL $\rightarrow$ LTL, converting a natural language string into a formal LTL expression.
    \item[]\label{prelim:verification}  \textbf{Verification:} Given a trace $\sigma$ or Kripke structure $K$, check whether a grounded LTL expression $g(\varphi)$ holds. For trace-based verification, construct a minimally satisfactory $K$ from $\sigma$. 
\end{enumerate}

\begin{table}[!b]
        \center
        \small
        \captionof{table}{Comparison of existing LTL benchmarks and \name{}. We report the number of unique words across all NL specifications and the number of unique LTL specifications. Additionally, we report support for lifting, grounding, and  verification.}

\begin{tabular}{lcccccc}
    \toprule
    Dataset
                                        & \# Words    
                                        & \# TL   
                                        & Lifting   & Translation & Grounding & Verification \\
    \midrule
    CW (\citet{CW_data})                & 184       & 37        &    $\sim$     & \checkmark & $\times$ & $\times$ \\
    GLTL (\citet{GLTL-data})            & 183       & 37        &    $\sim$     & \checkmark & $\times$  & $\times$ \\
    Navi (\citet{Navi_data})            & 131       & 6414      &    $\times$   & \checkmark & $\sim$ & $\times$ \\
    Conformal~(\citet{conformalLTL_data}) & 439     & 212       &    $\sim$     & $\checkmark$ & $\sim$ & $\times$ \\
    \midrule
    \name{} Warehouse                   & 1028      & 5991      &  \checkmark   & \checkmark & \checkmark & \checkmark \\
    \name{} Traffic Light               & 217       & 6196      &  \checkmark   & \checkmark & \checkmark & \checkmark \\
    \name{} Search and Rescue           & 245       & 5425      &  \checkmark   & \checkmark & \checkmark & \checkmark \\
   \textcolor{black}{\name{} Kitchen Assistant}           & \textcolor{black}{376}       & \textcolor{black}{7385}     &  \checkmark   & \checkmark & \checkmark & \checkmark \\

    \bottomrule
\end{tabular}

        \label{tab:qual_data_comp}
\end{table}

\subsection{Existing Benchmark Datasets}
In this section, we review existing benchmarks for NL-to-TL translation. We compare these corpora in terms of linguistic and logical complexity, and support for evaluation of different framework modules in Table \ref{tab:qual_data_comp}. We measure the complexity using the number of unique words appearing in natural language specifications (\#Words), as well as the number of unique temporal logic expressions (\#TL). In terms of modules, we report if a dataset has support for evaluation of lifting, grounding, and verification. We also provide examples from existing datasets in Appendix \ref{appendix:existing_data_examples}. 



In Table \ref{tab:qual_data_comp}, we observe that the older datasets \textbf{Cleanup World (CW)} \citep{CW_data} and \textbf{GLTL} \citep{GLTL-data} from the pre-LLM era have limited complexity both in terms of unique words and temporal logic expressions. While they support evaluation of translation, the lifting data is not explicitly given, the APs do not vary in their form to any meaningful degree, and they can be lexically identified in both the NL and TL elements of each entry (\texttt{"green room"} $\leftrightarrow G$, \texttt{"blue room"} $\leftrightarrow B$, etc.). The \textbf{Navi.} corpus, introduced by \citep{Navi_data}, couples NL commands with LTL formulas in a grid world. As Table~\ref{tab:qual_data_comp} shows, Navi exhibits a substantial increase in logical complexity, with 6,414 unique formulas and support for partial grounding. Its 221 unique APs make it a strong test of translation and lexical robustness, though this improvement comes at the cost of well-defined grounding rules: the corpus does not specify formal APs, providing instead POS-tagged natural language representations.  As reflected in Table~\ref{tab:qual_data_comp}, the \textbf{Conformal} \citep{conformalLTL_data} dataset introduces 439 unique words and 212 formulas with explicit grounding, but its scale is modest at 1,000 examples. In contrast, \name{} provides a testbed suited to holistic evaluation across lifting, translation, grounding, and verification. We provide a more detailed quantitative comparison between these datasets and \name{} in Section \ref{subsec:datasets}.



\section{The \fullname{}}
\label{sec:method}
In the following subsections, we first introduce \emph{Grounded Scenario Configuration}, which formalizes the world model by defining types, targets, and actions that ensure well-typed logical atoms. We then describe our \emph{Data Synthesis} pipeline, which instantiates expert-crafted NL–LTL templates with scenario-specific atoms to produce paired sentences, formulas, and traces. Next, we present the \emph{Metrics} used to evaluate each stage of the NL-to-LTL pipeline, and finally, we detail the \emph{Datasets} generated from three scenario definitions, highlighting their unique challenges and properties.


\subsection{Grounded Scenario Configuration}
\label{sec:scenarios}
To formalize how natural language specifications map onto executable logical structures, we distinguish three interconnected components: \textbf{types}, \textbf{targets}, and \textbf{actions}. \emph{Types} serve as abstract categories that describe what kinds of objects or entities an action can take as input (e.g., a location, an item, or a threat). \emph{Targets} are the grounded instantiations of these action--type combinations, where abstract slots are filled with concrete constants. \emph{Actions} are verbs that capture the capabilities of the agent; each action comes with a signature that specifies the expected types of its arguments. Together, this hierarchy ensures that linguistic expressions can be systematically mapped into well-typed logical atoms: types constrain argument structure, actions define the permissible predicates, and targets bind them to domain-specific instances. Each dataset is parameterized by a \emph{scenario}—a small, declarative world model that provides:

 \textbf{Types} $t\in\mathcal{T}$: denotes the sort of parameters accepted by an action (e.g.\  \texttt{item} or \texttt{location}).

 \textbf{Targets} $\mathcal{L}$: Specific instances of typed arguments, (e.g.\ an argument \texttt{apple} of type \texttt{item}, or an argument \texttt{loading\_dock} of type \texttt{location}).
 
\textbf{Actions} $\mathcal{A}_{\mathrm{args}}$: verbs the agent may perform, which may have one or more targets, (e.g.\ \texttt{\underline{idle}()} has no targets, \texttt{\underline{deliver}(apple, loading\_dock)} takes two---\texttt{item} and \texttt{location}).


 


\subsection{Data Synthesis}
\label{sec:synthesis}

To produce our datasets, we began with the 36 expert-crafted lifted NL-LTL pairs of the nl2spec benchmark \citep{nl2spec}, and we added 7 new ones of our own (provided in Appendix \ref{app:new_templates}). 
We then transformed these 43 examples into templates to support diverse NL–LTL synthesis.
Finally, for each NL–LTL example, we crafted one pair of traces---one satisfying and one violating. 

Each dataset entry includes a tuple of these three artifacts,
\[
  \bigl\{\,
    \underbrace{\mathrm{sentence},\,\mathrm{lifted \ sentence\}}}_{\text{NL (raw \& lifted)}},\;
    \underbrace{\varphi_G,\,\varphi_L}_{\text{LTL (grounded \& lifted)}},\;
    \underbrace{\sigma_{good} \models \varphi_G,\,\sigma_{bad}\not\models \varphi_G}_{\text{Traces (holds \& $\neg$ holds)}}
  \bigr\},
\]
and is algorithmically constructed with the following steps:
\begin{enumerate}
  \item \textbf{Template selection.} Uniformly choose a lifted template. Each template has an arity that determines how many atomic propositions must be instantiated. 

  \item \textbf{Atom sampling.} For each argument slot in the template, draw a unique atomic proposition by randomly selecting actions and arguments from the scenario’s $\mathcal{A}_t$ and $\mathcal{L}$. Let $k$ denote the total number of sampled atoms. Fill the LTL skeleton with these $k$ atoms to obtain the \emph{grounded} formula $\varphi_G$, and replace each atom by $\mathsf{prop}_i$ to obtain the \emph{lifted} formula $\varphi_L$.

  \item \textbf{NL realization.} Fill the template pattern with each atom’s surface form (including articles/prepositions), apply morphological fixes (gerunds, capitalization), and record token‐level spans. Emit both the free‐form sentence and its \texttt{grounded\_sentence} with explicit $\mathsf{prop}_i$ placeholders.

  \item \textbf{Trace filling.} Apply the template’s trace patterns to the list $[\mathsf{prop}_1,\dots,\mathsf{prop}_k]$, yielding one positive trace (satisfies $\varphi_G$) and one negative trace (violates $\varphi_G$).

\end{enumerate}
This rich annotation supports four independent evaluation axes, displayed in Figure \ref{fig:metrix}.

\subsection{Metrics}
\label{subsec:metrics}

\begin{figure}[!t]
    \centering
    \includegraphics[width=1.0\linewidth]{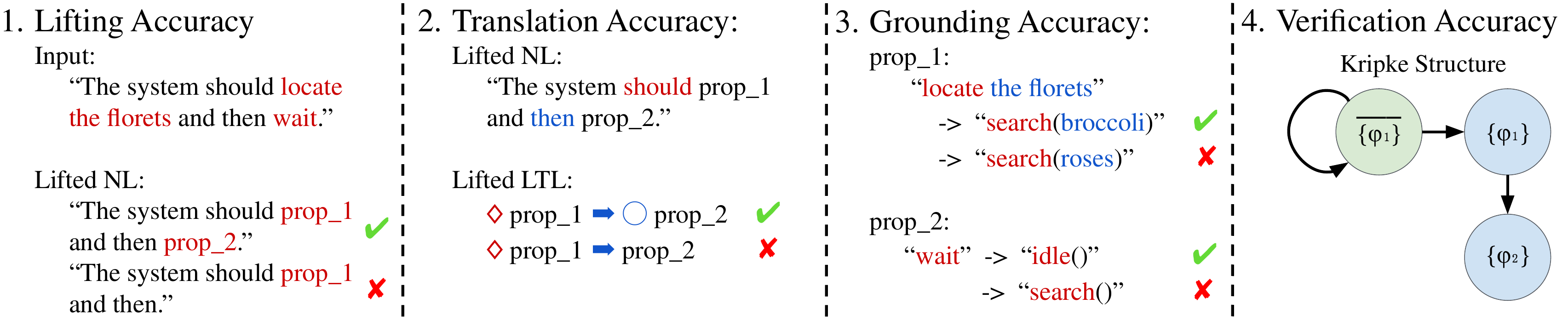}
    \caption{Overview of an isolated evaluation of each individual component. Lifting accuracy measures accuracy of predicted natural language AP spans, grounding accuracy measures the performance on mapping AP spans to world state conditions, translation accuracy measures the performance on NL-LTL translation on the token-level, and verification accuracy is an approach to measuring whether a grounded LTL expression holds on a trace.}
    \label{fig:metrix}
    
\end{figure}

In this section, we introduce four complementary metrics that capture performance at different levels of the NL-to-TL pipeline, which is illustrated in Figure~\ref{fig:metrix}. Lifting accuracy measures the identification of atomic proposition spans in natural language, grounding accuracy evaluates their mapping to world state conditions, translation accuracy assesses logical equivalence between predicted and reference formulas, and verification accuracy checks whether predicted formulas satisfy or violate traces as expected. Together, these metrics provide a comprehensive view of system performance.

\textbf{Lifting accuracy.}  
  For each token $\mathbb{S}_i$ in a sentence, the system predicts a label $\hat{\lambda}(\mathbb{S}_i) \in \{0,1,\dots,k\}$, where $0$ denotes background and $n$ denotes membership in $\pi_n$.  
  \[
  \text{LiftAcc} \;=\; \frac{1}{|\mathbb{S}|} \sum_{i=1}^{|\mathbb{S}|} \bigl[ \hat{\lambda}(\mathbb{S}_i) = \lambda(\mathbb{S}_i) \bigr].
  \]
  This measures the token-level classification accuracy of mapping substrings to atomic propositions.

\textbf{Translation accuracy.}  
  Given a natural language specification $s$, the system produces a predicted TL formula $\hat{\varphi}$. Translation accuracy is an exact match between the predicted and reference formulas:
  \[
  \text{TransAcc} \;=\; \bigl[ \hat{\varphi} \equiv \varphi \bigr],
  \]
  where $\equiv$ denotes logical equivalence. When working with lifted NL, the target is $\varphi_L$; for grounded NL, the target is $\varphi_G$. 

\textbf{Grounding accuracy.}  
  Let $\{\texttt{prop}_1,\dots,\texttt{prop}_k\}$ be lifted placeholders and $g_{\mathcal{S}}$ the gold grounding function. The system predicts $\hat{g}_{\mathcal{S}}$.  
  \[
  \text{GroundAcc} \;=\; \frac{1}{k} \sum_{j=1}^{k} \bigl[ \hat{g}_{\mathcal{S}}(\texttt{prop}_j) = g_{\mathcal{S}}(\texttt{prop}_j) \bigr].
  \]
  This measures how well predicted atoms match their reference predicates and arguments.

\textbf{Verification accuracy.}  
  For each dataset entry, two traces are provided: a positive trace $\sigma_{good}$ (satisfies $\varphi_G$) and a negative trace $\sigma_{bad}$ (violates $\varphi_G$). Given a predicted grounded formula $\hat{\varphi}_G$, verification checks whether the satisfaction relation holds:
  \[
  \text{VerifAcc} \;=\; \frac{1}{2} \Big(
      [ \sigma_{good} \models \hat{\varphi}_G ] \;+\;
      [ \sigma_{bad} \not\models \hat{\varphi}_G ]
  \Big).
  \]


\subsection{Datasets}
\label{subsec:datasets}
We construct \textcolor{black}{four} scenario definitions accompanied by action and target references, namely a \textcolor{black}{Kitchen Assistant}, Traffic Light, Search \& Rescue, and Warehouse scenario. The details are provided in Appendix~\ref{appendix:scenario_defs}. Using our proposed data synthesis, we generate \textcolor{black}{four} new datasets for training and evaluation. 
Each of our \textcolor{black}{four} datasets is designed to highlight distinct challenges for NL-to-LTL translation: the \emph{Traffic Light Control} scenario is intended to balance action and argument grounding challenges, including a large library of ``street name'' arguments, but a smaller set of actions; the \emph{Search-and-Rescue} scenario emphasizes multi-step temporal dependencies and deliberately includes ambiguous actions such as ``avoid'' and ``communicate'' to stress-test the system’s ability to distinguish between natural language verbs and temporal operators; \textcolor{black}{the \emph{Kitchen Assistant} scenario includes intentionally ambiguous action and argument references in order to stress both action and argument grounding simultaneously}; and the \emph{Warehouse} scenario introduces high semantic and linguistic variability by incorporating all 80 COCO object classes, making grounding especially complex. In this section, we use an entry from the \emph{Warehouse} dataset as an example to illustrate the structure and properties of our data; additional examples from the other scenarios are provided in Appendix~\ref{appendix:few_shot}.


\begin{table}[!t]
    \centering
    \caption{Comparison of NL–LTL datasets. We report the total number of entries (Size), the total number of unique TL entries, and the total number of unique APs appearing in the TL entries. $^\dagger$Note that these datasets do not explicitly provide quantities of actions and arguments, and these are estimated by the authors. 
    } 
    \label{tab:dataset-stats} 
        \begin{tabular}{lrrrrr} 
    \toprule 
    Dataset & \multicolumn{1}{c}{Size} & \multicolumn{1}{c}{Unique TL} & \multicolumn{1}{c}{\# APs} & \multicolumn{1}{c}{\# Actions} & \multicolumn{1}{c}{\# Args} \\ 
    \midrule 
    GLTL \citet{GLTL-data}$^\dagger$              & 11,109 &   37 &    4 & 1    &   4 \\
    CW   \citet{CW_data}$^\dagger$                  &  3,371 &   37 &    4 & 1    &   4 \\ Conformal \citet{conformalLTL_data}$^\dagger$   &  1,000 &  212 &  239 & 4    & 235 \\ 
    Navi  \citet{Navi_data}$^\dagger$               &  7,474 &6,414 &  221 & --   &  26 \\ 
    \midrule 
    \textcolor{black}{Kitchen Assistant     [\name]}       &  \textcolor{black}{10,000} & \textcolor{black}{7,385} & \textcolor{black}{172} & \textcolor{black}{9}    & \textcolor{black}{36} \\ 

    Search-and-rescue [\name]   &  7,304 &5,425 &  220 & 7    &  44 \\ 
    Traffic-light   [\name]     &  7,319 &6,196 &5,046 & 4    & 175 \\ 
    Warehouse      [\name]      &  7,457 &5,991 &5,074 & 5    & 82 \\ 
    \bottomrule 
\end{tabular}  
\end{table}

\paragraph{Warehouse.}
Our \emph{Warehouse} dataset simulates a realistic warehouse retrieval scenario, explicitly designed for scalability and complexity in grounding tasks. Warehouse is our most distinct dataset with its inclusion of all 80 COCO \citep{lin2014microsoft} object classes, significantly enriching the semantic and linguistic complexity and variation of atomic propositions. 
As with each of our datasets, all entries include LTL formulas with explicit grounding and alignment at token-level granularity, as well as verified positive ("good") and negative ("bad") execution traces for robust validation. 

\textbf{Example:}
\begin{itemize}
    \item \textbf{Sentence:} ``At every moment, at least one of drop off the long chair to the loading dock, wait, or look for the glass for alcoholic beverage holds.'' 
    \item \textbf{Lifted Sentence:} ``At every moment, at least one of \texttt{prop\_1}, \texttt{prop\_2} or \texttt{prop\_3} holds.''
    \item \textbf{Grounded LTL Formula:} \texttt{globally( deliver(bench, loading\_dock) or idle() or search(wine\_glass))}
    \item \textbf{APs:} \texttt{prop\_1} = ``drop off long chair to loading dock'', \texttt{prop\_2} = ``wait'', \texttt{prop\_3} = ``look for glass for alcoholic beverage''
    \item \textbf{Positive Trace:} [\texttt{deliver(bench, loading\_dock)}], [\texttt{idle()}], [\texttt{search(wine\_glass)}]
    \item \textbf{Negative Trace:} [\texttt{idle()}], [\texttt{idle()}], [\texttt{search(wine\_glass)}, \texttt{deliver(bench, loading\_dock)}]
\end{itemize}

\section{Experimental Results}
\label{sec:experiments}
In this section, we present the results of multiple evaluations of NL-to-LTL translation frameworks and components. In Section \ref{subsec:lifting_eval}, we measure the performance of common natural language lifting approaches, evaluated on four existing datasets in addition to \textcolor{black}{three of the four} datasets we present in \name{}. In Section \ref{subsec:translation_eval}, we evaluate three SOTA NL-to-LTL frameworks on lifted NL to lifted TL translation. Note here, that measuring lifted translation performance on existing datasets is particularly difficult, as they present varying degrees of clarity in their lifted natural language elements. In both translation evaluations, we use the pyModelChecking library \citep{pymodelcheck_code} to determine logical equivalence. The CW \citep{CW_data}, GLTL \citep{GLTL-data}, and Navi \citep{Navi_data} datasets have been processed to include lifted natural language components by \citep{NL2TL}, and we perform similar processing of the Conformal dataset \citep{conformalLTL_data} to include it in our evaluation. In Section \ref{subsec:grounding_eval} we develop and evaluate two grounding baselines on our three datasets. In Section \ref{subsec:endtoend_eval}, we assemble the best results from the three individual evaluations to perform the first end-to-end translation evaluation. In Section \ref{subsec:verification_eval} we perform our novel verification evaluation over the example traces of our dataset.

\subsection{Lifting Evaluation}
\label{subsec:lifting_eval}

First, we evaluate four language models on the natural language lifting task. The LLM-based approaches each use the lifting prompt template from the NL2TL framework \citep{NL2TL}, which includes few-shot ground-truth NL to lifted NL examples from each of the datasets. 
The input to both models is a natural language sentence and we compare the prediction made by the model against the ground-truth lifted natural language using the lifting accuracy metric defined in Section \ref{subsec:metrics}.  We present the results in Table \ref{tab:lifting_eval} where we see the linguistic complexity of our datasets is highlighted in the accuracies, as even the best scoring model (GPT-4.1) reduces in performance on our new datasets. 
This performance drop is even more significant on the lower-cost, smaller GPT models. This indicates our success in increasing evaluation complexity.  

\begin{table}[!ht]
    \centering
    \caption{Comparison of lifting approaches.}
        
\begin{tabular}{lccccccc}
\toprule
& \multicolumn{7}{c}{ Mean LiftAcc (\%)}\\
         & & & & & \multicolumn{1}{c}{S\&R} & \multicolumn{1}{c}{TL} & \multicolumn{1}{c}{WH} \\
        Model & \multicolumn{1}{c}{GLTL} & \multicolumn{1}{c}{CW} & \multicolumn{1}{c}{CF} & \multicolumn{1}{c}{Navi} & (ours) & (ours) & (ours) \\
\midrule
GPT-3.5-turbo    & 81.6 & 78.6 & 76.8 & 71.0 & 65.3 & 59.4 & 67.9 \\
GPT-4o-mini      & 84.9 & 82.3 & 85.6 & 81.0 & 66.7 & 63.1 & 68.9 \\
GPT-4.1-mini     & 97.7 & 95.9 & 96.1 & 97.1 & 94.4 & 96.6 & 93.1 \\
\bottomrule
\end{tabular}

    \label{tab:lifting_eval}
\end{table}

\subsection{Lifted Translation Evaluation}
\label{subsec:translation_eval}

Next, we evaluate the lifted translation capabilities of the three NL-to-LTL frameworks---nl2spec, NL2LTL, and NL2TL. In order to analyze the performance of their lifted translation abilities, the ground-truth lifted NL specification is given to the translation model, and the resulting lifted LTL translation is compared against the ground-truth lifted LTL. The formula for the translation accuracy metric is given in  Section~\ref{subsec:metrics}.
We present these results in Table \ref{tab:lifted_translation_eval}.
Here, we see that lifted translation can be very successful with both out-of-the-box LLM prompting (nl2spec) and with fine-tuned seq2seq models. However, as we have noted, we will see in end-to-end evaluation that this is an overconfident estimation of translation performance as grounding is not considered. 

\begin{table}[!ht]
    \centering
    \caption{Comparison of four frameworks on the Lifted NL to Lifted TL translation task. Note that we provide  ground-truth lifted NL specifications.}
    \resizebox{\textwidth}{!}{
        
  \begin{tabular}{llrrrrrrr}
    \toprule
    \multicolumn{1}{c}{ } &  \multicolumn{1}{c}{ } & \multicolumn{7}{c}{TransAcc (\%)}  \\
        &  & & & & & \multicolumn{1}{c}{S\&R} & \multicolumn{1}{c}{TL} & \multicolumn{1}{c}{WH} \\
        \multicolumn{1}{c}{Framework} & \multicolumn{1}{c}{Model} & \multicolumn{1}{c}{GLTL} & \multicolumn{1}{c}{CW} & \multicolumn{1}{c}{CF } & \multicolumn{1}{c}{Navi } & (ours) & (ours) & (ours) \\
    \midrule
    & GPT-3.5-turbo & 37.9 & 48.1 & 18.3 & 9.9 & 11.9 & 13.2 & 13.8 \\
       NL2LTL \citep{NL2LTL} & GPT-4o-mini   & 38.6 & 55.4 & 23.6 & 10.4  &12.3 & 13.9 & 12.5 \\
       & GPT-4.1-mini  & 51.7 & 64.6  & 42.1      & 39.7 & 41.6 & 40.0 & 37.4 \\
       \midrule

& GPT-3.5-turbo & 44.4 & 40.9 & 35.2 & 50.3 & 51.1 & 46.3 & 50.2 \\
   
   nl2spec \citep{nl2spec} & GPT-4o-mini   &  77.3 &  80.1 & 73.5 & 69.7 & 74.9 & 75.8 & 74.2\\
          & GPT-4.1-mini  & 89.8 & 92.9 & 78.3 & 81.5 & 89.1 & 91.6 & 88.4 \\
    
\midrule
  NL2TL \citep{NL2TL}, Lang2LTL &         t5-base                &  99.9 & 99.9        & 94.9  & 99.7 & 100.0          & 100.0       & 100.0  \\
\bottomrule
  \end{tabular}
    }
    \label{tab:lifted_translation_eval}
\end{table}

\subsection{Grounding Evaluation}
\label{subsec:grounding_eval}

In this section, we present the results obtained from our evaluation of our baseline grounding framework, applied to the ground truth lifted TL from our three \name{} datasets. We use two prompting strategies (described in Appendix \ref{appendix:prompts}) applied to three GPT models to provide a broad evaluation of current grounding capabilities. Our first prompting baseline---\textit{few-shot}---is composed of a brief description of the task at hand, accompanied by nine few-shot examples of correct (sentence, lifted sentence, AP-dictionary) tuples from \textit{all three scenarios} (as opposed to individual scenarios). The next strategy is the \textit{scenario} baseline prompt which includes the full scenario configuration file, as well as three few-shot examples from the dataset. \textcolor{black}{Our final grounding baseline employs the same scenario-specific few-shot examples as the previous approach, with the addition of specific instructions to include intermediate reasoning steps used to arrive at the answer. We use GPT-4o for this chain-of-thought approach in order to evaluate the performance of reasoning-capable models on this task. All four models are instructed to format their final answer in JSON format.} To measure grounding accuracy, we parse the resulting AP-dictionary predictions and compare them with our ground-truth knowledge of the AP-dictionary in each entry. Our metrics are per-AP and per-AP-dictionary accuracy. Per-AP accuracy is calculated by recording the total number of correctly grounded APs divided by the total number of APs in the test set, and per-AP-dictionary accuracy is calculated by recording the total number of completely correct AP-dictionaries, divided by the size of the test set. These results are presented in Table \ref{tab:grounding_eval}. 

Our evaluation of the two grounding baselines reveals that even advanced LLMs struggle to accurately ground lifted APs into a concrete world state space - even when the parameters of this state space are provided, as is done in the \emph{scenario} baseline. We observe that even though the \emph{scenario} baseline achieves lower performance on most benchmarks and settings, it beats the \emph{few-shot} baseline on our Warehouse scenario when comparing the more powerful reasoning models. As noted in Section \ref{sec:method}, the Warehouse scenario is specifically designed to stress-test \emph{grounding and lifting}. We conclude that the provision of the world state space in the \emph{scenario} baseline includes information that aids reasoning models in determining which world state conditions are referred to in the lifted APs, but the overall performance of these baselines on the grounding task remains notably lower than other tasks involved in verifiable NL-to-LTL translation.

\begin{table}[!ht]
    \centering
    \caption{Comparison of Grounding approaches. This table displays binary accuracy between predicted AP Grounding and known AP dictionary. LLM Baseline uses 9 few–shot sentence + lifted sentence + AP dict examples from every dataset; “Scenario” includes the scenario definition in the prompt and 3 examples from only that dataset. Note that Lang2LTL grounds using cosine similarity between reference and canonical AP embeddings.}
    \resizebox{\textwidth}{!}{
        
\begin{tabular}{lcccc|ccc}
\toprule
   && \multicolumn{3}{c}{Accuracy (\% of APs)} & \multicolumn{3}{c}{Accuracy (\% of AP Dictionaries)}\\
Prompt&Model            & S\&R      & Traffic Light  & Warehouse  & S\&R    & Traffic Light  & Warehouse  \\
\midrule
&GPT-3.5-turbo             & 56.9  & 69.5  & 18.3  & 34.2   & 51.4   & 7.4   \\
\emph{Few-shot General}&GPT-4o-mini                   & 82.3  & 66.5  & 18.4  & 68.6   & 48.4   & 7.0   \\
&GPT-4.1-mini             & 77.3  & 67.4  & 23.8  & 60.4   & 45.8   & 7.8   \\
\midrule
&GPT-3.5-turbo   & 76.7  & 37.3  & 13.6  & 63.6   & 20.8   & 5.0   \\
\textit{Few-shot Scenario}&GPT-4o-mini        & 66.7  & 44.8       & 23.6  & 44.8   & 16.8       & 9.2   \\
&GPT-4.1-mini   & 68.6  & 27.9  & 34.4  & 45.2   & 15.4   & 13.0  \\
\midrule
\textcolor{black}{\textit{Few-shot Chain-of-Thought}} &\textcolor{black}{GPT-4o} & \textcolor{black}{94.8}  & \textcolor{black}{85.9}  & \textcolor{black}{69.9}  & \textcolor{black}{94.1}   & \textcolor{black}{81.5}   & \textcolor{black}{61.4}  \\

\midrule
Lang2LTL \citep{Lang2LTL} & \textit{N/A} & 77.6 &86.2&61.8&59.0&73.6&38.8\\
\bottomrule
\end{tabular}

    }
    \label{tab:grounding_eval}
\end{table}

\subsection{End-to-End Translation Evaluation}
\label{subsec:endtoend_eval}

Now, we perform and end-to-end evaluation which considers the accumulation of the three individual translation steps. For all three frameworks, we select the best-performing component (model) from each of the individual evaluations (lifting, grounding, and translation) to assemble an end-to-end translation framework which factors in the combined performance of all the translation steps. We see in Table \ref{tab:endtoend}, that as a result of the poor grounding results of all current approaches, the high performance of the lifting and lifted translation steps is diminished, resulting in a poor overall semantic accuracy of the final translation. Our datasets show that even the best performing model (NL2TL) does not approach real-world performance needs, inciting the need for NL-to-TL translation approaches which consider a concrete world state space. 

\begin{table}[!h]
    \centering
    \caption{End-to-end evaluation of all three SOTA frameworks using the best lifting, translation, and grounding components. We report the binary accuracy of the resulting LTL.}
      \begin{tabular}{lrrr}
    \toprule
        \multicolumn{1}{c}{}      & \multicolumn{3}{c}{Accuracy (\%)} \\

    \multicolumn{1}{c}{Framework}      & \multicolumn{1}{c}{S\&R} & \multicolumn{1}{c}{Traffic Light} & \multicolumn{1}{c}{Warehouse} \\
    \midrule
       NL2LTL \citep{NL2LTL} & 35.4 & 38.4 & 26.2 \\
   
       nl2spec \citep{nl2spec} & 34.8 & 33.6 & 29.6 \\

      NL2TL \citep{NL2TL} & 54.4& 60.1 & 46.2 \\
    Lang2LTL \citep{Lang2LTL} & 58.5 & 72.1 & 37.9\\
              \bottomrule
  \end{tabular}  

    \label{tab:endtoend}
\end{table}

\subsection{Verification Evaluation}
Finally, we present the results of our experiments on the verification of LTL outputs from each of the three NL-to-LTL translation frameworks that we compare. We use the outputs from our lifted translation evaluation (Table \ref{tab:lifted_translation_eval}) to isolate the verification metric from the lifting task, and apply our LLM-baseline grounding frameworks. In Table \ref{tab:verification_eval}, out results demonstrate that even frameworks exhibiting accurate lifted NL to lifted TL translation suffer a notable decline in performance when grounding relies on systems similar to our LLM baselines. Furthermore, this evaluation supports the use of trace satisfaction in place of ground-truth LTL comparison as a metric for grounded translation accuracy, because the example traces encode the minimum specifications of correctly grounded and translated LTL. In future frameworks, example traces could be used as part of a feedback loop to grounding and translation components. 
\label{subsec:verification_eval}

\begin{table}[!ht]
    \centering
    \caption{Performance (binary accuracy) on S\&R, Traffic Light, and Warehouse, broken down into satisfied holding traces, satisfied not-holding traces, and both. All three frameworks are evaluated on both grounding strategies using their top-scoring lifted translation model.}
    \resizebox{\textwidth}{!}{








\begin{tabular}{ll *{3}{rrr}}
    \toprule
       &  & \multicolumn{3}{c}{S\&R} & \multicolumn{3}{c}{Traffic Light} & \multicolumn{3}{c}{Warehouse} \\
    \cmidrule(lr){3-5} \cmidrule(lr){6-8} \cmidrule(lr){9-11}
     Framework& Grounding Strategy  & Sat & Unsat & Both & Sat & Unsat & Both & Sat & Unsat & Both \\
    \midrule
    \multirow{3}{*}{NL2LTL \citep{NL2LTL}}
      & \emph{Few-shot General}  & 61.6  & 61.4  & 35.4  & 64.6  & 60.2  & 38.4  & 52.4  & 58.6  & 26.2 \\     
      & \emph{Few-shot Scenario}  &  1.06 & 32.0  &  7.4  & 61.8  & 59.2  & 36.6  & 12.4  & 36.2  &  9.8 \\
    \midrule
    \multirow{3}{*}{nl2spec \citep{nl2spec}}
      & \emph{Few-shot General}  & 47.4  & 48.0  & 34.8  & 47.2  & 46.0  & 33.6  & 46.0  & 44.2  & 29.6 \\
      & \emph{Few-shot Scenario}  & 34.0  & 36.4  & 21.0  & 40.2  & 41.8  & 28.2  & 32.0  & 34.6  & 19.0 \\
    \midrule
    \multirow{3}{*}{NL2TL \citep{NL2TL}}
      & \emph{Few-shot General}  & 75.0  & 79.4  & 54.4  & 80.2  & 80.6  & 60.8  & 71.4  & 74.8  & 46.2 \\
      & \emph{Few-shot Scenario}  & 27.5  & 50.8  & 22.1  & 72.6  & 76.3  & 54.5  & 33.3  & 52.4  & 23.5 \\
      \midrule
    Lang2LTL \citep{Lang2LTL}
      & Embedding  & 43.3 & 61.9  & 39.3  & 44.7  & 63.0  & 41.3  & 21.6  & 40.1  & 16.6 \\
    \bottomrule
\end{tabular}%

    }
    \label{tab:verification_eval}
\end{table}





\section{Conclusion}
\label{sec:conclusion}
We present the \fullname{}. \name{} is a suite of three new NL-to-LTL translation datasets that include the standard natural language and LTL pairs, supplemented with lifted natural language, lifted LTL, and trace examples. These additional features provide a method for the isolated training and evaluation of individual NL-to-LTL translation framework components. The provision of trace examples in \name{} introduces the possibility of a new type of input that is plausible in real-world translation frameworks, but unrepresented in current corpora. We acknowledge that the datasets included in the \name{} suite are generated using a finite number of linguistic and logical templates, populated by diverse synthetic natural language APs. \name{} reveals significant weaknesses in what were previously ironclad NL-to-LTL translation frameworks. Among these weakness are: the reliance on accurately lifted NL inputs for translation, lack of accurate grounding components, and lack of example trace inputs in current approaches. We envision our contribution will encourage exploration of diverse methods for grounded NL-to-LTL translation, beyond the use of LLMs.


\clearpage

\bibliography{iclr2026_conference}

@InProceedings{Navi_data,
  title = 	 {Learning a natural-language to LTL executable semantic parser for grounded robotics},
  author =       {Wang, Christopher and Ross, Candace and Kuo, Yen-Ling and Katz, Boris and Barbu, Andrei},
  booktitle = 	 {Proceedings of the 2020 Conference on Robot Learning},
  pages = 	 {1706--1718},
  year = 	 {2021},
  editor = 	 {Kober, Jens and Ramos, Fabio and Tomlin, Claire},
  volume = 	 {155},
  series = 	 {Proceedings of Machine Learning Research},
  month = 	 {16--18 Nov},
  publisher =    {PMLR},
  url = 	 {https://proceedings.mlr.press/v155/wang21g.html},
}

@article{GLTL-data,
  title={Sequence-to-Sequence Language Grounding of Non-Markovian Task Specifications},
  author={Nakul Gopalan and Dilip Arumugam and Lawson L. S. Wong and Stefanie Tellex},
  journal={Robotics: Science and Systems XIV},
  year={2018},
  url={https://api.semanticscholar.org/CorpusID:46994194}
}

@inproceedings{CW_data,
  title={Grounding English Commands to Reward Functions},
  author={James MacGlashan and Monica Babes-Vroman and Marie desJardins and Michael L. Littman and Smaranda Muresan and S. Squire and Stefanie Tellex and Dilip Arumugam and Lei Yang},
  booktitle={Robotics: Science and Systems},
  year={2015},
  url={https://api.semanticscholar.org/CorpusID:1709515}
}

@inproceedings{Natural2CTL_data,
author = {Zrelli, Rim and Amaral Misson, Henrique and Ben Attia, Maroua and Gohring de Magalh\~{a}es, Felipe and Shabah, Abdo and Nicolescu, Gabriela},
title = {Natural2CTL: A Dataset for Natural Language Requirements and Their CTL Formal Equivalents},
year = {2024},
isbn = {978-3-031-57326-2},
publisher = {Springer-Verlag},
address = {Berlin, Heidelberg},
url = {https://doi.org/10.1007/978-3-031-57327-9_13},
doi = {10.1007/978-3-031-57327-9_13},
booktitle = {Requirements Engineering: Foundation for Software Quality: 30th International Working Conference, REFSQ 2024, Winterthur, Switzerland, April 8–11, 2024, Proceedings},
pages = {205–216},
numpages = {12},
location = {Winterthur, Switzerland}
}

@misc{conformalLTL_data,
      title={ConformalNL2LTL: Translating Natural Language Instructions into Temporal Logic Formulas with Conformal Correctness Guarantees}, 
      author={Jun Wang and David Smith Sundarsingh and Jyotirmoy V. Deshmukh and Yiannis Kantaros},
      year={2025},
      eprint={2504.21022},
      archivePrefix={arXiv},
      primaryClass={cs.CL},
      url={https://arxiv.org/abs/2504.21022}, 
}

@misc{circuits_data,
      title={DeepSTL -- From English Requirements to Signal Temporal Logic}, 
      author={Jie He and Ezio Bartocci and Dejan Ničković and Haris Isakovic and Radu Grosu},
      year={2022},
      eprint={2109.10294},
      archivePrefix={arXiv},
      primaryClass={cs.CL},
      url={https://arxiv.org/abs/2109.10294}, 
}

@inproceedings{NL2TL,
Author = {Yongchao Chen and Rujul Gandhi and Yang Zhang and Chuchu Fan},
Title = {NL2TL: Transforming Natural Languages to Temporal Logics using Large Language Models},
Year = {2023},
Eprint = {arXiv:2305.07766},
booktitle = {Proceedings of the 2023 Conference on Empirical Methods in Natural
  Language Processing},
}

@INPROCEEDINGS{STL,
  author={Madsen, Curtis and Vaidyanathan, Prashant and Sadraddini, Sadra and Vasile, Cristian-Ioan and DeLateur, Nicholas A. and Weiss, Ron and Densmore, Douglas and Belta, Calin},
  booktitle={2018 IEEE Conference on Decision and Control (CDC)}, 
  title={Metrics for Signal Temporal Logic Formulae}, 
  year={2018},
  volume={},
  number={},
  pages={1542-1547},
  doi={10.1109/CDC.2018.8619541}}

@inproceedings{nl2spec,
    author = {Cosler, Matthias and Hahn, Christopher and Mendoza, Daniel and Schmitt, Frederik and Trippel, Caroline},
    title = {nl2spec: Interactively Translating Unstructured Natural Language to Temporal Logics with Large Language Models},
    year = {2023},
    isbn = {978-3-031-37702-0},
    publisher = {Springer-Verlag},
    address = {Berlin, Heidelberg},
    url = {https://doi.org/10.1007/978-3-031-37703-7_18},
    doi = {10.1007/978-3-031-37703-7_18},
    booktitle = {Computer Aided Verification: 35th International Conference, CAV 2023, Paris, France, July 17–22, 2023, Proceedings, Part II},
    pages = {383–396},
    numpages = {14},
    location = {Paris, France}
}

@inproceedings{NL2LTL,
  title={NL2LTL - a Python Package for Converting Natural Language (NL) Instructions to Linear Temporal Logic (LTL) Formulas},
  author={Francesco Fuggitti and Tathagata Chakraborti},
  booktitle={AAAI Conference on Artificial Intelligence},
  year={2023},
  url={https://api.semanticscholar.org/CorpusID:259726762}
}

@book{SAS_textbook_PoCPS,
author = {Alur, Rajeev},
title = {Principles of Cyber-Physical Systems},
year = {2015},
isbn = {0262029111},
publisher = {The MIT Press},

}

@article{SurveyTemporal,
author = {Savas Konur},
title = {A survey on temporal logics for specifying and verifying real-time systems},
publisher = {Front. Comput. Sci.},
year = {2013},
journal = {Frontiers of Computer Science},
volume = {7},
number = {3},
eid = {370},
numpages = {33},
pages = {370},
url = {https://journal.hep.com.cn/fcs/EN/abstract/article_4956.shtml},
doi = {10.1007/s11704-013-2195-2}
}

@INPROCEEDINGS{LTL,
  author={Zhu, Weijun},
  booktitle={2021 IEEE 4th Advanced Information Management, Communicates, Electronic and Automation Control Conference (IMCEC)}, 
  title={Big Data on Linear Temporal Logic Formulas}, 
  year={2021},
  volume={4},
  number={},
  pages={544-547},
  doi={10.1109/IMCEC51613.2021.9482368}}

@article{RobotLanguage,
   author = "Tellex, Stefanie and Gopalan, Nakul and Kress-Gazit, Hadas and Matuszek, Cynthia",
   title = "Robots That Use Language", 
   journal= "Annual Review of Control, Robotics, and Autonomous Systems",
   year = "2020",
   volume = "3",
   number = "Volume 3, 2020",
   pages = "25-55",
   doi = "https://doi.org/10.1146/annurev-control-101119-071628",
   url = "https://www.annualreviews.org/content/journals/10.1146/annurev-control-101119-071628",
   publisher = "Annual Reviews",
   issn = "2573-5144",
   type = "Journal Article"
  }

@misc{Spec2,
author = {Raman, Vasumathi and Lignos, Constantine and Finucane, Cameron and Lee, Kenton and Marcus, Mitch and Kress-Gazit, Hadas},
year = {2013},
month = {06},
pages = {},
title = {Sorry Dave, I'm Afraid I Can't Do That: Explaining Unachievable Robot Tasks Using Natural Language},
doi = {10.15607/RSS.2013.IX.023}
}

@article{yin2024formal,
  title={Formal synthesis of controllers for safety-critical autonomous systems: Developments and challenges},
  author={Yin, Xiang and Gao, Bingzhao and Yu, Xiao},
  journal={Annual Reviews in Control},
  volume={57},
  pages={100940},
  year={2024},
  publisher={Elsevier}
}

@article{cardoso2021review,
  title={A review of verification and validation for space autonomous systems},
  author={Cardoso, Rafael C and Kourtis, Georgios and Dennis, Louise A and Dixon, Clare and Farrell, Marie and Fisher, Michael and Webster, Matt},
  journal={Current Robotics Reports},
  volume={2},
  number={3},
  pages={273--283},
  year={2021},
  publisher={Springer}
}

@article{thistle1986control,
  title={Control problems in a temporal logic framework},
  author={Thistle, JG and Wonham, WM},
  journal={International Journal of Control},
  volume={44},
  number={4},
  pages={943--976},
  year={1986},
  publisher={Taylor \& Francis}
}

@article{watson2005autonomous,
  title={Autonomous systems},
  author={Watson, David P and Scheidt, David H},
  journal={Johns Hopkins APL technical digest},
  volume={26},
  number={4},
  pages={368--376},
  year={2005}
}

@misc{post2018fastlexicallyconstraineddecoding,
      title={Fast Lexically Constrained Decoding with Dynamic Beam Allocation for Neural Machine Translation}, 
      author={Matt Post and David Vilar},
      year={2018},
      eprint={1804.06609},
      archivePrefix={arXiv},
      primaryClass={cs.CL},
      url={https://arxiv.org/abs/1804.06609}, 
}

@misc{geng2024grammarconstraineddecodingstructurednlp,
      title={Grammar-Constrained Decoding for Structured NLP Tasks without Finetuning}, 
      author={Saibo Geng and Martin Josifoski and Maxime Peyrard and Robert West},
      year={2024},
      eprint={2305.13971},
      archivePrefix={arXiv},
      primaryClass={cs.CL},
      url={https://arxiv.org/abs/2305.13971}, 
}

@inproceedings{xu2024learning,
  title={Learning from Failures: Translation of Natural Language Requirements into Linear Temporal Logic with Large Language Models},
  author={Xu, Yilongfei and Feng, Jincao and Miao, Weikai},
  booktitle={2024 IEEE 24th International Conference on Software Quality, Reliability and Security (QRS)},
  pages={204--215},
  year={2024},
  organization={IEEE}
}

@article{hahn2022formal,
  title={Formal specifications from natural language},
  author={Hahn, Christopher and Schmitt, Frederik and Tillman, Julia J and Metzger, Niklas and Siber, Julian and Finkbeiner, Bernd},
  journal={arXiv preprint arXiv:2206.01962},
  year={2022}
}

@inproceedings{pan2023data,
  title={Data-efficient learning of natural language to linear temporal logic translators for robot task specification},
  author={Pan, Jiayi and Chou, Glen and Berenson, Dmitry},
  booktitle={2023 IEEE International Conference on Robotics and Automation (ICRA)},
  pages={11554--11561},
  year={2023},
  organization={IEEE}
}

@inproceedings{hsiung2022generalizing,
  title={Generalizing to new domains by mapping natural language to lifted LTL},
  author={Hsiung, Eric and Mehta, Hiloni and Chu, Junchi and Liu, Xinyu and Patel, Roma and Tellex, Stefanie and Konidaris, George},
  booktitle={2022 International Conference on Robotics and Automation (ICRA)},
  pages={3624--3630},
  year={2022},
  organization={IEEE}
}

@article{bellini2000temporal,
  title={Temporal logics for real-time system specification},
  author={Bellini, Pierfrancesco and Mattolini, Riccardo and Nesi, Paolo},
  journal={ACM Computing Surveys (CSUR)},
  volume={32},
  number={1},
  pages={12--42},
  year={2000},
  publisher={ACM New York, NY, USA}
}

@article{veizaga2021systematically,
  title={On systematically building a controlled natural language for functional requirements},
  author={Veizaga, Alvaro and Alferez, Mauricio and Torre, Damiano and Sabetzadeh, Mehrdad and Briand, Lionel},
  journal={Empirical Software Engineering},
  volume={26},
  number={4},
  pages={79},
  year={2021},
  publisher={Springer}
}

@article{lafi2021eliciting,
  title={Eliciting requirements from stakeholders' responses using natural language processing},
  author={Lafi, Mohammed and Hawashin, Bilal and AlZu'bi, Shadi},
  journal={Computer Modeling In Engineering \& Sciences},
  volume={127},
  number={1},
  pages={99--116},
  year={2021},
  publisher={Tech Science Press}
}

@mastersthesis{lamar2009linguistic,
  title={Linguistic analysis of natural language engineering requirements},
  author={Lamar, Carl},
  year={2009},
  school={Clemson University}
}

@misc{pymodelcheck_code,
    author = {Alberto Casagrande},
    title = {pyModelChecking Code Repository},
    year= {2024},
  note = {Available at \url{https://github.com/albertocasagrande/pyModelChecking}}
}

@inproceedings{lin2014microsoft,
  title={Microsoft coco: Common objects in context},
  author={Lin, Tsung-Yi and Maire, Michael and Belongie, Serge and Hays, James and Perona, Pietro and Ramanan, Deva and Doll{\'a}r, Piotr and Zitnick, C Lawrence},
  booktitle={Computer vision--ECCV 2014: 13th European conference, zurich, Switzerland, September 6-12, 2014, proceedings, part v 13},
  pages={740--755},
  year={2014},
  organization={Springer}
}

@InProceedings{Lang2LTL,
  title = 	 {Grounding Complex Natural Language Commands for Temporal Tasks in Unseen Environments},
  author =       {Liu, Jason Xinyu and Yang, Ziyi and Idrees, Ifrah and Liang, Sam and Schornstein, Benjamin and Tellex, Stefanie and Shah, Ankit},
  booktitle = 	 {Proceedings of The 7th Conference on Robot Learning},
  pages = 	 {1084--1110},
  year = 	 {2023},
  editor = 	 {Tan, Jie and Toussaint, Marc and Darvish, Kourosh},
  volume = 	 {229},
  series = 	 {Proceedings of Machine Learning Research},
  month = 	 {06--09 Nov},
  publisher =    {PMLR},
  url = 	 {https://proceedings.mlr.press/v229/liu23d.html}
}
\bibliographystyle{iclr2026_conference}

\clearpage

\appendix

\section{Appendix}
\label{app}
In this appendix, we present a detailed overview of linear temporal logic in \ref{appendix:ltl}, a discussion of verification via Kripke structures in \ref{appendix:modelcheck}, a quantitative comparison of our \name{} dataset against existing datasets as well as examples from those datasets in \ref{appendix:existing_data_examples}, our developed prompts for the baseline grounding approaches in \ref{appendix:prompts}, the configuration files for our three scenarios in \ref{appendix:scenario_defs}, and finally our estimated compute resource usage and our external code and license information in \ref{appendix:compute}.

\subsection{Linear Temporal Logic}
\label{appendix:ltl}
Linear temporal logic (LTL) is a modal extension of classical propositional logic that enables reasoning about how truths evolve over a discrete, linear timeline~\citep{LTL}. Formulas in LTL are interpreted over infinite sequences (or “traces”) of states
$$
\sigma = s_0, s_1, s_2, \ldots,
$$
where each state $s_i$ (which has a set of conditions) specifies which atomic propositions $\pi^\mu$ hold true at time $i$. This framework makes it possible to specify and verify both safety properties (e.g., “nothing bad ever happens”) and liveness properties (e.g., “something good eventually happens”), and it underpins many model-checking techniques for reactive systems.

The syntax of LTL is given by the following grammar:
\begin{align*}
    \varphi ::={} & \pi 
    \;\mid\; \neg \varphi 
    \;\mid\; \varphi_1 \wedge \varphi_2 
    \;\mid\; \varphi_1 \vee \varphi_2 
    \;\mid\; \varphi_1 \Rightarrow \varphi_2 \nonumber\\
    &\;\mid\; \bigcirc \varphi 
    \;\mid\; \diamondsuit \varphi 
    \;\mid\; \Box \varphi 
    \;\mid\; \varphi_1 \,\cup\, \varphi_2
    \label{eq:ltl-syntax}
\end{align*}
where
$\pi$ ranges over a finite set of atomic propositions;  $\neg$, $\wedge$, $\vee$, and $\Rightarrow$ are the standard Boolean connectives; $\bigcirc$ (\texttt{next}) asserts that its operand holds in the immediately following state; $\diamondsuit$ (\texttt{eventually}) asserts that its operand holds at some point in the future;  $\Box$ (\texttt{always}) asserts that its operand holds at every future state;  $\varphi_1 \cup \varphi_2$ (\texttt{until}) asserts that $\varphi_1$ continuously holds until $\varphi_2$ becomes true. Formally, we write 
$\sigma, i \models \varphi$
to mean “formula $\varphi$ holds at position $i$ in trace $\sigma$.”  For example:
\begin{align*}
  \sigma, i \models \varphi_1 \cup \varphi_2 
    &\quad\text{iff}\quad \exists k \ge i:\; \sigma, k \models \varphi_2
      \;\wedge\;\forall j \in [i,k):\;\sigma, j \models \varphi_1.
\end{align*}

Although our focus is on discrete-time LTL, many of these ideas carry over to related formalisms such as signal temporal logic (STL) for continuous-time, real-valued signals~\citep{STL}.

\clearpage
\subsection{Verification via Kripke Structures and Fluents}
\label{appendix:modelcheck}
Verification of LTL specifications is typically conducted using a Kripke structure, which is a formal transition system comprising states, transitions, and labels indicating which atomic propositions hold true in each state. Formally, a Kripke structure is defined as a tuple $M = (S, S\_0, R, L)$, where:

\begin{itemize}
    \item $S$ is a finite set of states,
    \item $S_0 \subseteq S$ is the set of initial states,
    \item $R \subseteq S \times S$ is the transition relation, specifying allowed state transitions,
    \item $L: S \rightarrow 2^{\mathit{AP}}$ is a labeling function mapping states to the sets of atomic propositions that are true in each state.
\end{itemize}

Verification involves checking whether every possible path through the Kripke structure satisfies the given LTL formula. For instance, safety properties such as \textit{“a collision never occurs”} require that no path through the structure contains a state labeled with the proposition \texttt{collision}. Conversely, liveness properties such as \textit{“a goal is eventually reached”} demand the existence of a future state in every valid path labeled with the proposition \texttt{goal}. Additionally, verification explicitly involves fluents—timestamped state variables that indicate when certain conditions or states become true. Each fluent captures both the state variable (atomic proposition) and the time step at which the transition into the corresponding state occurs. Formally, a fluent can be represented as a tuple $(\pi^\mu, t)$, indicating that proposition $\pi^\mu$ becomes true at time step $t$ due to a state transition within the Kripke structure. Fluents bridge the gap between high-level temporal specifications and lower-level state transitions, facilitating practical model checking and control synthesis in robot control systems.

\clearpage
\subsection{VLTL-Bench LTL Expression Statistics}

\begin{table}[h]
\centering
\begin{tabular}{lccc}
\toprule
\textbf{Token / Operator} & \textbf{Search \& Rescue} & \textbf{Traffic Light} & \textbf{Warehouse} \\
\midrule

and            & 5536  & 5508  & 5297  \\
double\_implies& 1104  & 1141  & 1091  \\
finally        & 3835  & 3842  & 3918  \\
globally       & 9899  & 9851  & 9820  \\
implies        & 5164  & 5295  & 5264  \\
next           & 12144 & 12304 & 11713 \\
not            & 6781  & 6724  & 6593  \\
or             & 3198  & 3229  & 3054  \\
prop\_1        & 14440 & 14466 & 14193 \\
prop\_2        & 7934  & 7964  & 7835  \\
prop\_3        & 3740  & 3831  & 3825  \\
until          & 1112  & 1088  & 1147  \\
\bottomrule
\end{tabular}
\caption{Operator splits and template breakdowns by domain.}
\end{table}
\subsection{VLTL-Bench New Templates}
\label{app:new_templates}
We then craft 7 of our own templates to fill perceived gaps in specification coverage. Of these templates, 4 entries include new lifted LTL halves (marked below with a *), and 3 include new lifted NL halves. 

\begin{table}[!ht]
    \centering
    \small
        \begin{tabular}{|l|p{9cm}|}
            \hline
            \textbf{NL} & \textbf{LTL} \\
            \hline
            \texttt{finally ( not prop\_1)} & ``eventually, avoid prop\_1'' \\
            \hline
            \texttt{globally ( not prop\_1)} & ``always avoid prop\_1''; ``prop\_1 must never occur'' \\
            \hline
            \texttt{next prop\_1} & ``at the next time step, prop\_1 holds'' \\
            \hline
            \texttt{prop\_1 until prop\_2} & ``prop\_1 must always hold at all times before prop\_2'' \\
            \hline
            \texttt{finally (prop\_1 and prop\_2)} & OLD: ``Eventually, both prop\_1 and prop\_2 will hold simultaneously'' \\ 
            & NEW: ``At some point, prop\_1 and prop\_2 will both hold at the same time.'' \\
            \hline
            \texttt{globally (prop\_1 and prop\_2)} & OLD: ``Both prop\_1 and prop\_2 hold at every step.'' \\ 
            & NEW: ``At all time steps, prop\_1 and prop\_2 both hold.'' \\
            \hline
            \texttt{finally (prop\_1 or prop\_2)} & OLD: ``eventually, either prop\_1 or prop\_2'' \\ 
            & NEW: ``either prop\_1 or prop\_2 will hold at some point in time.'' \\
            \hline
        \end{tabular}
    \caption{Examples of NL–LTL mappings. OLD/NEW entries show updated phrasing.}
\end{table}

\subsection{Existing Datasets}
\label{appendix:existing_data_examples}

\paragraph{Cleanup World (CW).}
\begin{itemize}
    \item Sentence: ``go to the blue room keep going and stop when you reach the green room''
    \item LTL Formula: ``finally(blue\_room and finally green\_room)'' 
    \item Grounded Sentence: ``go to the prop\_1 keep going and stop when you reach the green prop\_2,'' 
    \item APs: prop\_1 = go to blue room, prop\_2 = go to green room.
\end{itemize}

\paragraph{GLTL.}
\begin{itemize}
    \item Sentence: ``enter the blue or red room and proceed until the green room''
    \item LTL Formula: ``finally((red\_room or blue\_room) and finally green\_room)'' 
    \item Grounded Sentence: ``enter the prop\_2 or prop\_1 and proceed until the green prop\_3,'' 
    \item APs: prop\_1 = go to red room, prop\_2 = go to blue room, prop\_3 = go to green room
\end{itemize}
\paragraph{Navi.}
\begin{itemize}
    \item Sentence: ``at some time get hold apple or whenever acquire pear''
    \item LTL Formula: ``finally(get\_hold\_v apple\_n or finally(acquire\_v pear\_n)'' 
    \item Grounded Sentence: ``at some time prop\_1 or whenever prop\_2'' 
    \item APs:  prop\_1 = get\_hold\_v apple\_n, prop\_2 = acquire\_v pear\_n
\end{itemize}
\paragraph{ConformalNL2LTL.}
\begin{itemize}
    \item Sentence: ``Stay in parking lot 4 until you reach car 5''
    \item LTL Formula: ``parking\_lot\_4 until car\_5'' 
    \item Grounded Sentence: ``Stay in prop\_1 until you reach prop\_2'' 
    \item APs: prop\_1 = go to parking lot 4, prop\_2 = go to car 5
\end{itemize}
\vspace{3pt}

\subsection{Grounding Prompts}
\label{appendix:prompts}

This section includes the few-shot examples used in our grounding prompt baselines. The \textit{few-shot} baselines uses all of the following in its prompt, while the \textit{scenario} baseline includes only the scenario specific few-shot examples combined with the scenario description, given in Appendix \ref{appendix:scenario_defs}

\textbf{Few-shot Prompt:}

\fbox{%
    \begin{minipage}{\dimexpr\linewidth-2\fboxsep-2\fboxrule}
{"role": "system", "content": "You are an LTL translation assistant, your goal is to return the desired prop\_dict, a dictionary that relates natural language atomic proposition/predicate references to their canonical/known representation in the scenario."},

{"role": "user",   "content": 
\\
Few-shot Examples:
\\
\{\texttt{examples from ALL domains, shown in appendix \ref{appendix:few_shot}, total of 9 examples}\} 

Now predict:

Sentence: \{\texttt{sentence}\}

Lifted: \{\texttt{lifted\_sentence}\}

Prop\_dict:  

}
    \end{minipage}%
}

\textbf{Scenario Prompt:}

\fbox{%
    \begin{minipage}{\dimexpr\linewidth-2\fboxsep-2\fboxrule}
{"role": "system", "content": "You are an LTL translation assistant, your goal is to return the desired prop\_dict, a dictionary that relates natural language atomic proposition/predicate references to their canonical/known representation in the scenario."},

{"role": "user",   "content": 
\\
Scenario Configuration:
\texttt{{scenario yaml, given in appendix \ref{appendix:scenario_defs}}}
\\
Few-shot Examples:

\{\texttt{examples from this specific scenario, shown in Appendix \ref{appendix:few_shot}}\}
\\
Now predict:

Sentence: \{\texttt{sentence}\}
\\
Lifted: \{\texttt{lifted\_sentence}\}
\\
Prop\_dict:  

}
    \end{minipage}%
}

\clearpage
\subsection{Few-shot Examples by Scenario}
\label{appendix:few_shot}

\textbf{Warehouse Examples}

\fbox{%
    \begin{minipage}{\dimexpr\linewidth-2\fboxsep-2\fboxrule}
Sentence: ["The system must eventually, avoid prop\_1"]\\
Lifted Sentence: ["The system must eventually, avoid prop\_1"]\\
prop\_dict: \{\\
  "prop\_1": \{\\
    "action\_canon": "deliver",\\
    "action\_ref":   "drop off",\\
    "args\_canon":   ["sandwich loading\_dock"],\\
    "args\_ref":     ["square food loading dock"]\\
  \}\\
\}

Sentence: ["Whenever prop\_1 holds, prop\_2 holds as well."]\\
Lifted Sentence: ["Whenever prop\_1 holds, prop\_2 holds as well."]\\
prop\_dict: \{\\
  "prop\_1": \{\\
    "action\_canon": "idle",\\
    "action\_ref":   "remain still",\\
    "args\_canon":   [],\\
    "args\_ref":     []\\
  \},\\
  "prop\_2": \{
    "action\_canon": "get\_help",\\
    "action\_ref":   "call for help",\\
    "args\_canon":   [],\\
    "args\_ref":     []\\
  \}\\
\}\\

Sentence: ["If prop\_2 holds, then in the next step prop\_3 persists until prop\_1 holds, or else prop\_3 holds forever."]\\
Lifted Sentence: ["If prop\_2 holds, then in the next step prop\_3 persists until prop\_1 holds, or else prop\_3 holds forever."]\\
prop\_dict: \{\\
  "prop\_1": \{\\
    "action\_canon": "pickup",\\
    "action\_ref":   "grab",\\
    "args\_canon":   ["hot\_dog"],\\
    "args\_ref":     ["bunned sausage"]\\
  \},\\
  "prop\_2": \{\\
    "action\_canon": "pickup",\\
    "action\_ref":   "grab",\\
    "args\_canon":   ["potted\_plant"],\\
    "args\_ref":     ["plant"]\\
  \},\\
  "prop\_3": \{
    "action\_canon": "search",\\
    "action\_ref":   "search for",\\
    "args\_canon":   ["cup"],\\
    "args\_ref":     ["beverage cup"]\\
  \}\\
\}

    \end{minipage}%
}
\clearpage
\textbf{Search and Rescue Examples}

\fbox{%
    \begin{minipage}{\dimexpr\linewidth-2\fboxsep-2\fboxrule}

Sentence: ["This controller must always avoid prop\_1"]\\
Lifted Sentence: ["This controller must always avoid prop\_1"]\\
prop\_dict: \{\\
  "prop\_1": \{\\
    "action\_canon": "record",\\
    "action\_ref":   "begin recording",\\
    "args\_canon":   ["fire\_source"],\\
    "args\_ref":     ["fire source"]\\
  \}\\
\}\\

Sentence: ["In this task, take a photo of flood, then return home."]\\
Lifted Sentence: ["In this task, prop\_1 then prop\_2"]\\
prop\_dict: \{\\
  "prop\_1": \{\\
    "action\_canon": "photo",\\
    "action\_ref":   "take a photo of",\\
    "args\_canon":   ["flood"],\\
    "args\_ref":     ["flood"]\\
  \},\\
  "prop\_2": \{\\
    "action\_canon": "go\_home",\\
    "action\_ref":   "return home",\\
    "args\_canon":   [],\\
    "args\_ref":     []\\
  \}\\
\}\\

Sentence: ["If every record flood is eventually followed by talking to the safe victim, then avoid the impending debris must occur infinitely often."]\\
Lifted Sentence: ["If every prop\_1 is eventually followed by prop\_2 then prop\_3 must occur infinitely often."]\\
prop\_dict: \{\\
  "prop\_1": \{\\
    "action\_canon": "record",\\
    "action\_ref":   "record",\\
    "args\_canon":   ["flood"],\\
    "args\_ref":     ["flood"]\\
  \},\\
  "prop\_2": \{\\
    "action\_canon": "communicate",\\
    "action\_ref":   "talk to",\\
    "args\_canon":   ["safe\_victim"],\\
    "args\_ref":     ["safe victim"]\\
  \},\\
  "prop\_3": \{\\
    "action\_canon": "avoid",\\
    "action\_ref":   "avoid",\\
    "args\_canon":   ["impending\_debris"],\\
    "args\_ref":     ["impending debris"]\\
  \}\\
\}\\

    \end{minipage}%
}

\clearpage
\textbf{Traffic Light Examples}

\fbox{%
    \begin{minipage}{\dimexpr\linewidth-2\fboxsep-2\fboxrule}

Sentence: ["You", "must", "eventually,", "avoid", "set", "east", "light", "yellow."]\\
Grounded: ["You", "must", "eventually,", "avoid", "prop\_1"]\\
prop\_dict: \{\\
  "prop\_1": \{\\
    "action\_canon": "change",
    "action\_ref":   "set",
    "args\_canon":   ["light\_east", "yellow"],
    "args\_ref":     ["east light", "yellow"]
  \}
\}

Sentence: ["Both", "change", "west", "light", "red", "and", "take", "a", "video", "of", "the", "car", "on", "southwest", "10th", "avenue", "hold", "at", "every", "step."]\\
Grounded: ["Both", "prop\_1", "and", "prop\_2", "hold", "at", "every", "step."]\\
prop\_dict: \{\\
  "prop\_1": \{\\
    "action\_canon": "change",\\
    "action\_ref":   "change",\\
    "args\_canon":   ["light\_west", "red"],\\
    "args\_ref":     ["west light", "red"]\\
  \},\\
  "prop\_2": \{\\
    "action\_canon": "record",\\
    "action\_ref":   "take a video of",\\
    "args\_canon":   ["car", "southwest\_10th\_avenue"],\\
    "args\_ref":     ["car", "southwest 10th avenue"]\\
 \}
\}

Sentence: ["If", "take", "a", "picture", "of", "the", "car", "on", "northwest", "6th", "street", "holds", "and", "set", "east", "light", "green", "holds", "next,", "then", "request", "assistance", "holds", "in", "the", "step", "after", "that."]\\
Grounded: ["If", "prop\_1", "holds", "and", "prop\_2", "holds", "next,", "then", "prop\_3", "holds", "in", "the", "step", "after", "that."]\\
prop\_dict: \{\\
  "prop\_1": \{\\
    "action\_canon": "photo",\\
    "action\_ref":   "take a picture of",\\
    "args\_canon":   ["car", "northwest\_6th\_street"],\\
    "args\_ref":     ["car", "northwest 6th street"]\\
  \},\\
  "prop\_2": \{\\
    "action\_canon": "change",\\
    "action\_ref":   "set",\\
    "args\_canon":   ["light\_east", "green"],\\
    "args\_ref":     ["east light", "green"]\\
  \},\\
  "prop\_3": \{\\
    "action\_canon": "get\_help",\\
    "action\_ref":   "request assistance",\\
    "args\_canon":   [],\\
    "args\_ref":     []\\
  \}\\
\}

    \end{minipage}%
}




\clearpage

\subsection{Scenario Configurations}
\label{appendix:scenario_defs}
In this section, we provide the scenario configuration files that are inserted into the grounding prompts and used for data generation.

\begin{figure}[!ht]
    \centering
    \includegraphics[width=.7\linewidth]{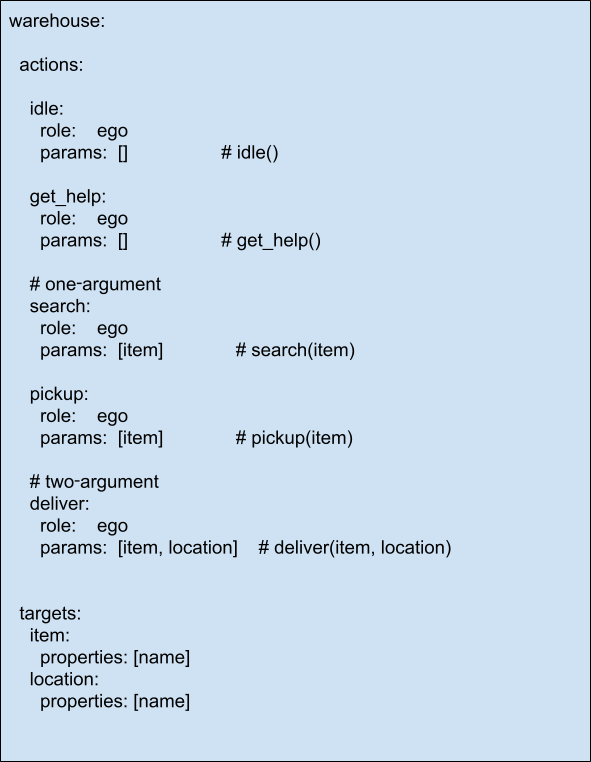}
    \caption{Warehouse Scenario Configuration file}
    \label{fig:Warehouse_Config}
\end{figure}

\begin{figure}[!ht]
    \centering
    \includegraphics[width=0.7\linewidth]{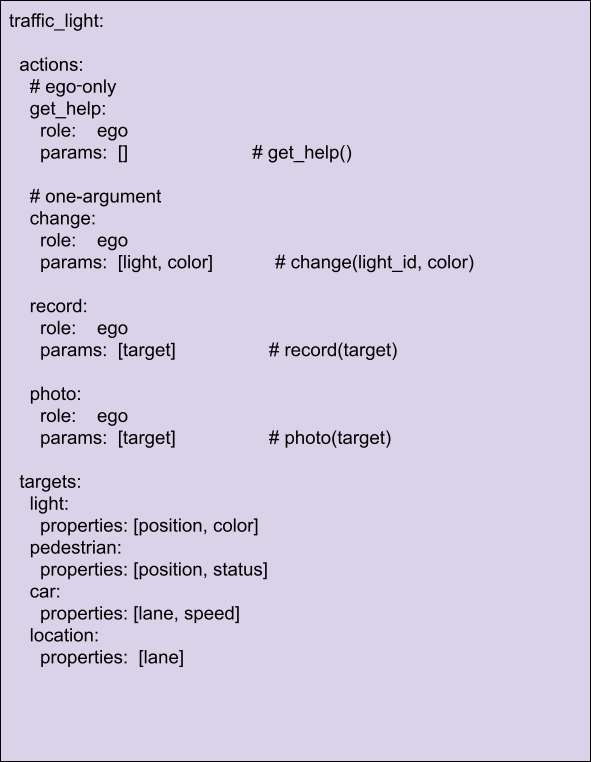}
    \caption{Traffic Light Scenario Configuration file}
    \label{fig:Traffic_Config}
\end{figure}

\begin{figure}[!ht]
    \centering
    \includegraphics[width=0.7\linewidth]{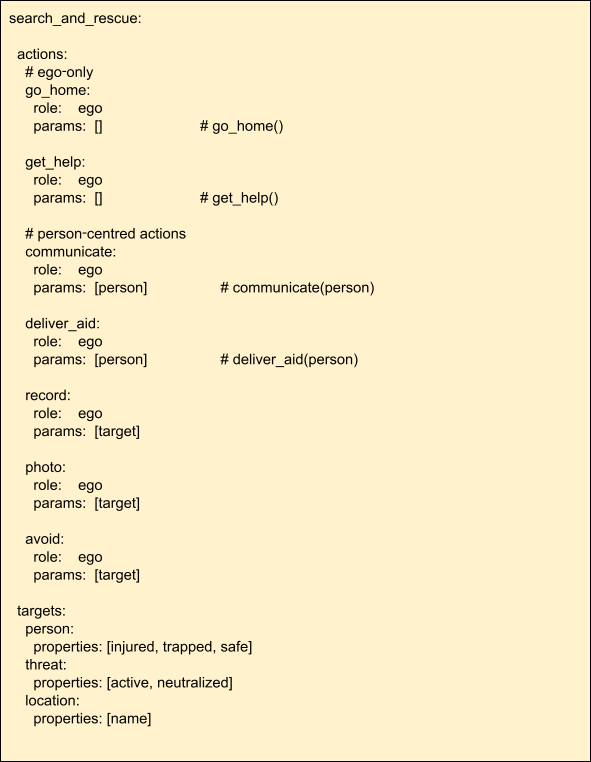}
    \caption{Search and Rescue Scenario Configuration file}
    \label{fig:SR_Config}
\end{figure}

\begin{figure}[!ht]
    \centering
    \includegraphics[width=0.85\linewidth]{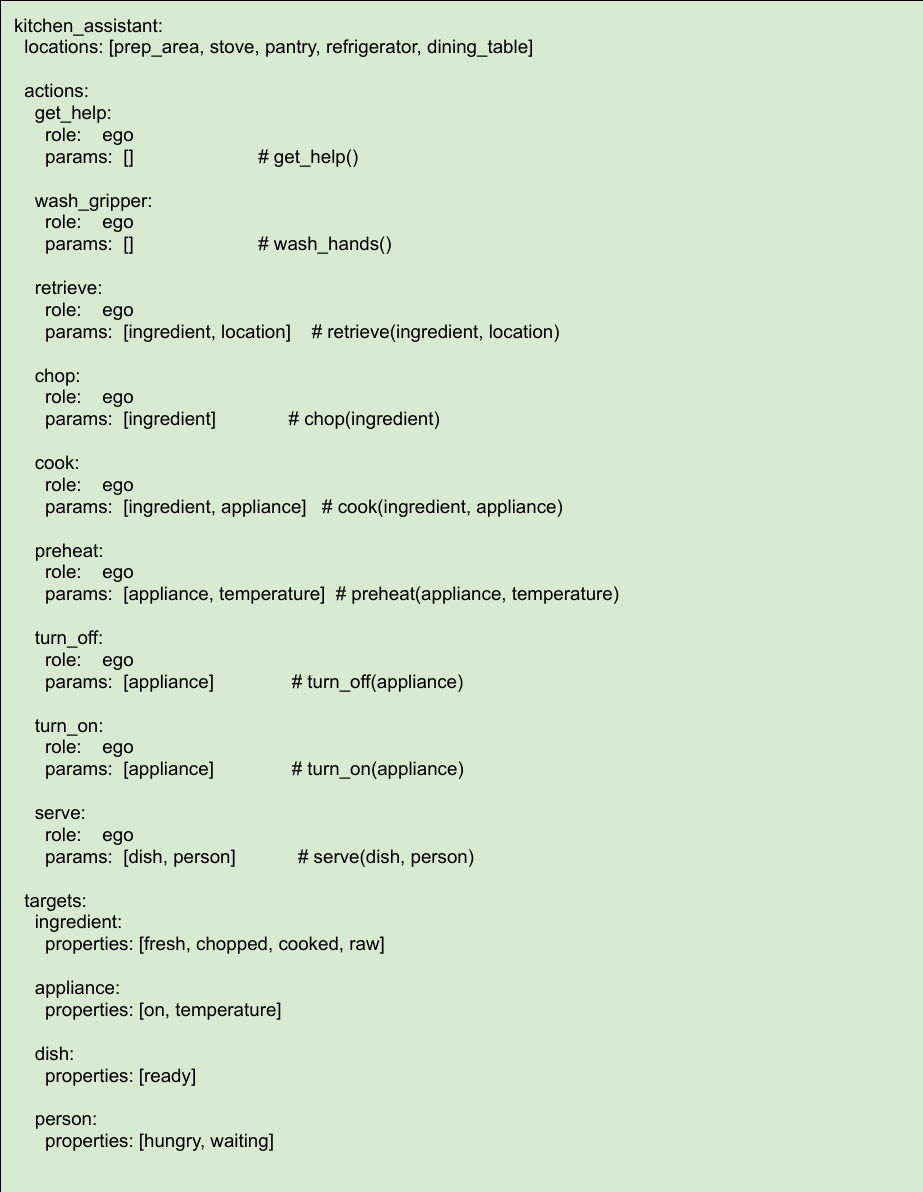}
    \caption{\textcolor{black}{Kitchen Assistant Scenario Configuration file}}
    \label{fig:KA_Config}
\end{figure}
\clearpage
\subsection{Compute Resources and External Code and License Information}
\label{appendix:compute}
All LLM inference was performed using the OpenAI API. Approximately \$30.00 in compute credits were used for our evaluations. The T5-base model used by NL2TL was trained and tested locally on a machine using an Nvidia GeForce RTX 4070Ti Super 16 GB GPU, an Intel i9 14900KF, and 64 GB of RAM. Training took approximately 40 minutes using a batch size of 16 and a learning rate of $2\text{e}^{-5}$ for 3 epochs. 

The nl2spec framework is released at \url{https://github.com/realChrisHahn2/nl2spec} under the MIT license, the NL2TL framework is released at \url{https://github.com/yongchao98/NL2TL?tab=readme-ov-file} with no attached license, the NL2LTL framework is released at \url{https://github.com/IBM/nl2ltl} under the MIT license, and the pyModelChecking library is released at \url{https://github.com/albertocasagrande/pyModelChecking} under the GNU General Public License.

\subsection{Large Language Model Disclosure}
During the preparation of this paper, the authors employed large language models (LLMs) as assistive tools for limited tasks including proof-reading, text summarization, and the discovery of related work. All substantive research contributions, analyses, and claims presented in this paper were conceived, developed, and verified by the authors. The authors maintain full ownership and responsibility for the content of the paper, including its technical correctness, originality, and scholarly contributions.
\end{document}